\documentclass[aps,preprintnumbers, superscriptaddress, amsmath]{revtex4}

\usepackage{graphicx}
\usepackage{dcolumn}
\usepackage{bm}
\begin{document}

\title{EoS's of different phases of  dense quark matter}

\author{E J Ferrer\\ \textit{Dept. of Engineering Science and Physics, College of Staten Island,
CUNY,\\ and CUNY-Graduate Center,
           New York 10314, USA}}

\begin{abstract}
Compact stars with significant high densities in their interiors can give rise to quark deconfined phases that can open a window for the study of strongly interacting dense nuclear matter. Recent observations on the mass of two pulsars, PSR J1614-2230 and PSR J0348+0432, have posed a great restriction on their composition, since their equations of state must be hard enough to support masses of about at least two solar masses. The onset of quarks tends to soften the equation of state, but 
due to their strong interactions, different phases can be realized with new parameters that affect the corresponding equations of state and ultimately the mass-radius relationships. In this paper I will review how the equation of state of dense quark matter is affected by the physical characteristics of the phases that can take place at different baryonic densities with and without the presence of a magnetic field, as well as their connection with the corresponding mass-radius relationship, which can be derived by using different models.
\end{abstract}

\maketitle

\section{Introduction}

The composition of neutron stars is still an open question. Propositions ranging from nuclear matter (possibly with hyperons and superfluid nucleons) to deconfined quark matter (in any of its possible phases as color superconducting or with inhomogeneous particle-hole pairing, etc.) are under exam. New data on masses and radii, transport properties, as well as the modeling of other phenomena as glitches, bursts episodes, etc., help to constraint the equation of state (EoS) corresponding to different matter phases that can prevail in their interior, but no firm conclusion has been drawn yet.
 In both, nuclear and quark matter descriptions, there still exist a few parameters to be adjusted, which should be further constrained by nuclear matter and heavy-ion collision data, so to be able to discriminate among the different star constituents,
 but their region of reliability at low temperature and large chemical potential poses a big challenge.
 
The proposition that quark stars could be a stable configuration for compact stellar objects has been around for almost 50 years \cite{Bodmer, Chin, Terazawa, Wit}. The idea is based on the fact that matter composed of up, down, and strange quarks, the so called strange quark matter (SQM), may have a lower energy per baryon number
than the nucleon, thus being absolutely stable.
More recent developments brought the idea that color superconductivity should be the favored state of deconfined quark matter,
since the Cooper pair condensation lower the total energy per baryon number of the system
\cite{Pairing1, Pairing2, Pairing3, German}. On this basis, many works have been done to study the characteristics of the EoS of the different phases of quark matter.

 In this regard, recent data \cite{Ozels1, Ozels2, Demorest}
have determined masses and/or radii for some compact objects with high precision (although some of these remain to be confirmed \cite{Lattimer}), rendering some information about the possible composition of these objects.
In this scenario, the recent observation for the pulsar J1614-2230, a binary system for which the mass of the neutron star was measured rather accurately through the Shapiro delay yielding M = 1.97$\pm$0.04 $M_{\odot}$ \cite{Demorest}, posed an important question about the existence of SQM. Of course the answer to this is directly related to the prevailing matter phase that can produce an EoS stiff enough to reach such large mass value. 
 
 In this paper, I am reviewing several results obtained in a set of papers that will be conveniently referred, where different approaches in describing the quark matter EoS were used. A main purpose will be to discuss some effects that can affect the stiffness of the EoS, so to determine if it is possible to reach the reported mass value M = 1.97$\pm$0.04 $M_{\odot}$. I will finish discussing the magnetic field effect on the EoS of quark matter in the color superconducting color-flavor-locked (CFL) phase, pointing out the new challenges that the presence of a magnetic field poses on the calculation of the M-R relationship.

\section{Bag vs NJL model in the CFL phase}

In the study of the EoS of the possible matter phases that could form neutron stars,  two frameworks have been mainly used: the MIT bag model and the Nambu-Jona-Lasinio (NJL) model. Both models are inspired in some of the fundamental properties of strongly interacting systems, although some aspects could be lacking.

The MIT bag model is a phenomenological model that was proposed to explain hadrons \cite{MIT}. In this model, the quarks are asymptotically free and confinement is provided through the bag constant $B$, which artificially constrains the quarks inside a finite region in space. 
In the case of bulk quark matter, the asymptotically-free phase of quarks will form a perturbative vacuum (inside a "bag"), which is immersed in the nonperturbative vacuum. The creation of the 'bag" costs free energy. Then,
in the energy density, the energy difference between the perturbative vacuum and the true one should be added. Essentially, that is the bag constant
characterizing a constant energy per unit volume associated to the region where the quarks live. From the point of view of the pressure,
$B$ can be interpreted as an inward pressure needed to confine the quarks into the bag. 

Let's consider now the difference between these two approaches in the case of SQM. For the free quark system, the thermodynamic potential is given in the MIT framework by

\begin{equation}\label{Omega}
\Omega=\sum_i\Omega_i+B,
\end{equation}
where
\begin{equation}\label{Omega1}
\Omega_i=\frac{\mu_i^4}{4\pi^2},
\end{equation}
with $i$ running for  quarks u, d, s and electrons, and $\mu_i$ being the corresponding
chemical potential for each particle $i$. 

When the color superconducting pairing is considered in the CFL phase, which is the most stable phase at high densities \cite{Alford04}, the thermodynamical potential is assumed to be the sum of the one corresponding to the unpaired state (\ref{Omega}), plus a gap ($\Delta$) depending term \cite{Rajagopal2001, Alford2001}, 
\begin{equation}
\Omega_{CFL}^{MIT}=\sum_i\Omega_i-\frac{3}{\pi^2}\mu^2\Delta_{CFL}^2+B\label{Omega_MIT}
\end{equation}
The term $\sum_i\Omega_i$ represents a fictitious non-paired
state proportional to the volume of the Fermi sphere in which all quarks have a common Fermi momentum and the extra
$\Delta$-depending term represents a surface contribution of the Fermi sphere that is associated with the Cooper pair biding energy. Here, as in (\ref{Omega}), $B$ represents the bag constant contribution.

In the context of the NJL model, the thermodynamic potential of the CFL phase is given by \cite{Paulucci},
\begin{eqnarray}
\Omega_{CFL}^{NJL} =-\frac{1}{4\pi^2}\int_0^\infty dp p^2 e^{-p^2/\Lambda^2}(16|
\epsilon|+16|\overline{\epsilon}|)+ \nonumber \\
-\frac{1}{4\pi^2}\int_0^\infty
dp p^2 e^{-p^2/\Lambda^2}(2|\epsilon'|+2|\overline{\epsilon'}|)+
\frac{3\Delta_{CFL}^2}{G}+B,
 \label{OmegaCFL}
\end{eqnarray}
where
\begin{eqnarray}
\varepsilon=\pm \sqrt{(p-\mu)^2+\Delta_{CFL}^2}, \quad
\overline{\varepsilon}=\pm \sqrt{(p+\mu)^2+\Delta_{CFL}^2},\nonumber
\\
\varepsilon'=\pm \sqrt{(p-\mu)^2+4\Delta_{CFL}^2,}\quad
 \overline{\varepsilon'}=\pm \sqrt{(p+\mu)^2+4\Delta_{CFL}^2},
 \label{6}
\end{eqnarray}
are the quasiparticles dispersion relations. See that in (\ref{6}) the particle masses are neglected at the high chemical potentials where the CFL phase is realized. In (\ref{OmegaCFL}) a smooth cutoff depending on  the effective-theory energy scale $\Lambda$ was introduced to guarantee continuous thermodynamical quantities.

The system EoS is then obtained from the thermodynamic potential as the relation between the energy density and the pressure, which for an isotropic system can be respectively obtained as
\begin{equation}  \label{Energy-Density}
\epsilon=\Omega-\mu \frac{\partial \Omega}{\partial \mu},
\end{equation}
\begin{equation}  \label{Pressure}
P=-\Omega
\end{equation}

On the other hand, the mass-radius (M-R) relationship is obtained by integrating the relativistic
equations for stellar structure, that is, the well-known Tolman-Oppenheimer-Volkoff (TOV) and mass continuity equations, which in  natural units, $c = G = 1$ are given by
\begin{eqnarray}
\frac{dM}{dR}&=&4\pi R^2\epsilon \label{TOV1}\\
\frac{dP}{dR}&=&-\frac{\epsilon M}{R^2}\Big(1+\frac{P}{\epsilon}\Big)\Big
(1+\frac{4\pi R^3P}{M}\Big)\Big(1-\frac{2M}{R}\Big)^{-1}\label{TOV2}
\end{eqnarray}
with $\epsilon$ and $P$ taken from (\ref{Energy-Density}) and (\ref{Pressure}) respectively.

\begin{figure}
\begin{center}
\includegraphics[width=0.48\textwidth]{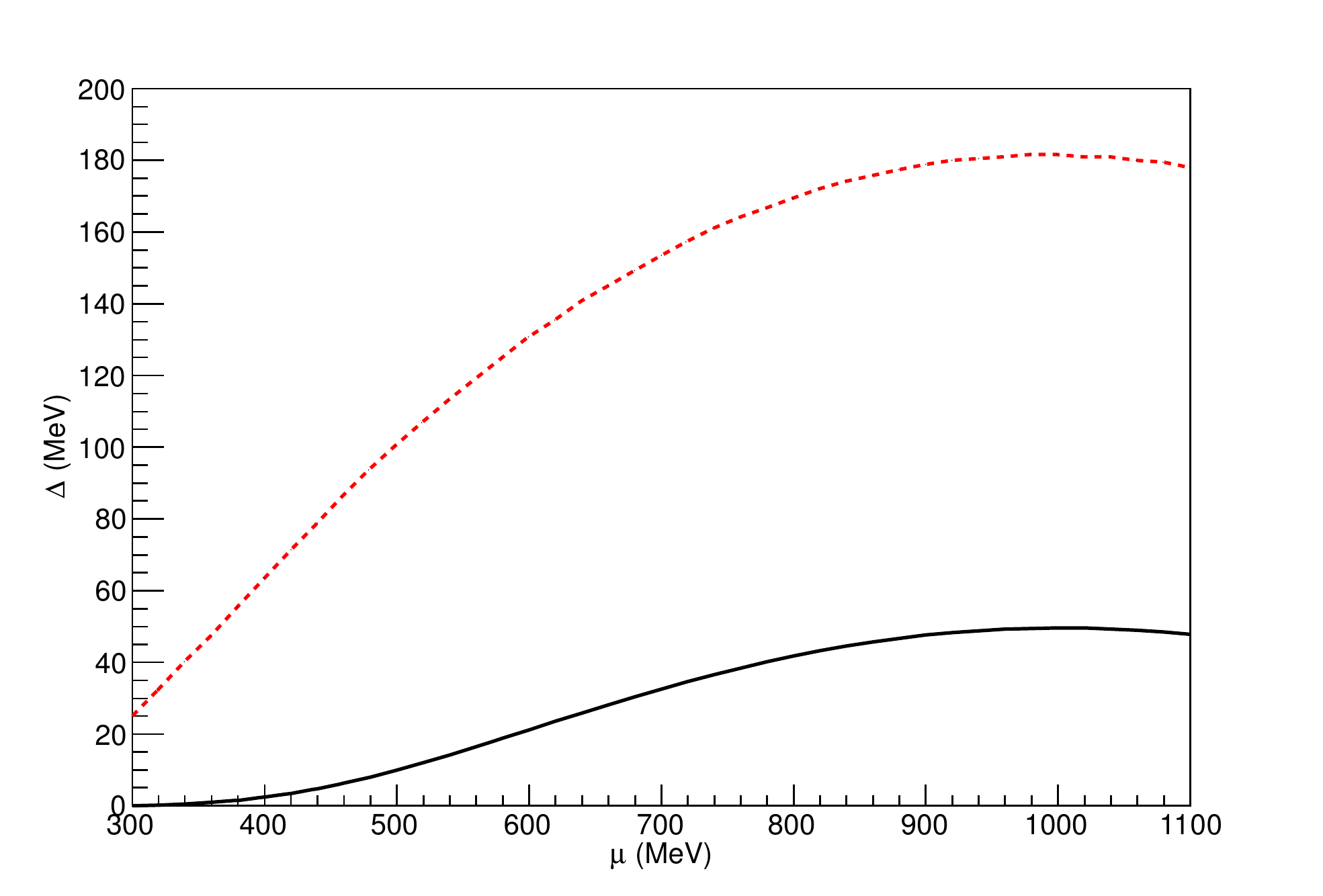}
\caption{\footnotesize Gap parameter, $\Delta$, vs baryonic chemical potential, $\mu$, for NJL model in the CFL phase for two coupling constant values G=4.32 GeV$^{-2}$ (full line) and G=7.10 GeV$^{-2}$ (dashed line).} \label{Gap}
\end{center}
\end{figure}

A fundamental difference between  models (\ref{Omega_MIT}) and (\ref{OmegaCFL}) is related to the way the pairing term $\Delta$ is implemented.
In the MIT bag model the value for the gap parameter is fixed by hand, hence it does not explicitly depend on changes in other parameters characterizing the system.
While in the NJL approach the pairing gap is self-consistently obtained through the gap equation
\begin{equation}  \label{Gap-Eq-1}
 \frac{\partial \Omega_{CFL}}{\partial\Delta_{CFL}}=0. 
 \end{equation} 
 In this way, as it was shown in Ref. \cite{Incera}, 
 $\Delta$ depends on the density and diquark coupling constant $G$, as can be seen in Fig \ref{Gap}. Thus, changing the coupling constant in this case from G = 4.32 to 7.10 GeV$^{-2}$, the corresponding gap parameters obtained at $\mu=500$ MeV change from $\Delta$ = 10 to 100 MeV, respectively. We point out that the value $G=7.10 GeV^{-2}$ is used with the only purpose of comparing the NJL results with the MIT ones, where the use of $\Delta= 100$ MeV is a common practice.

\begin{figure}
\begin{center}
\includegraphics[width=0.49\textwidth]{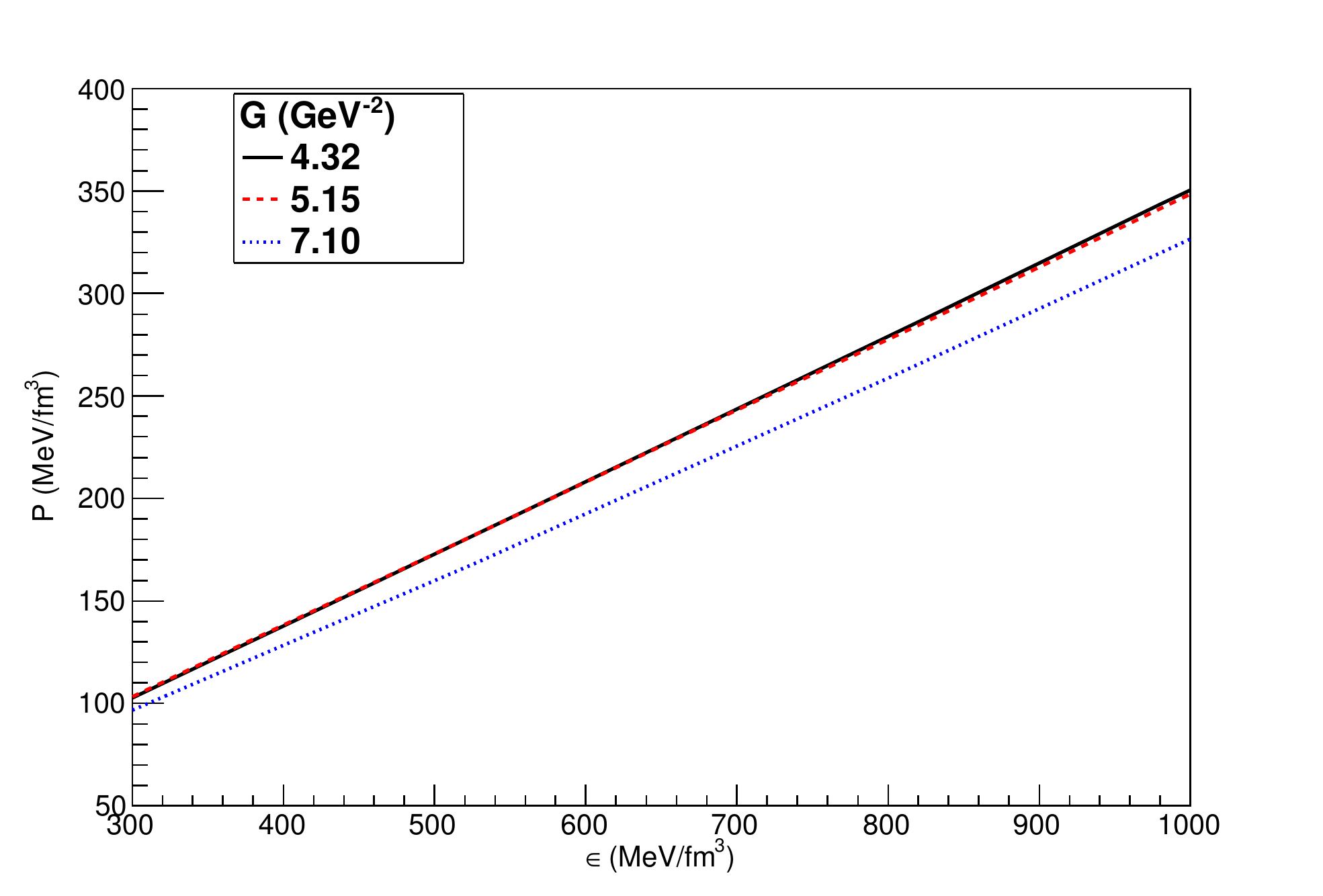}
\includegraphics[width=0.49\textwidth]{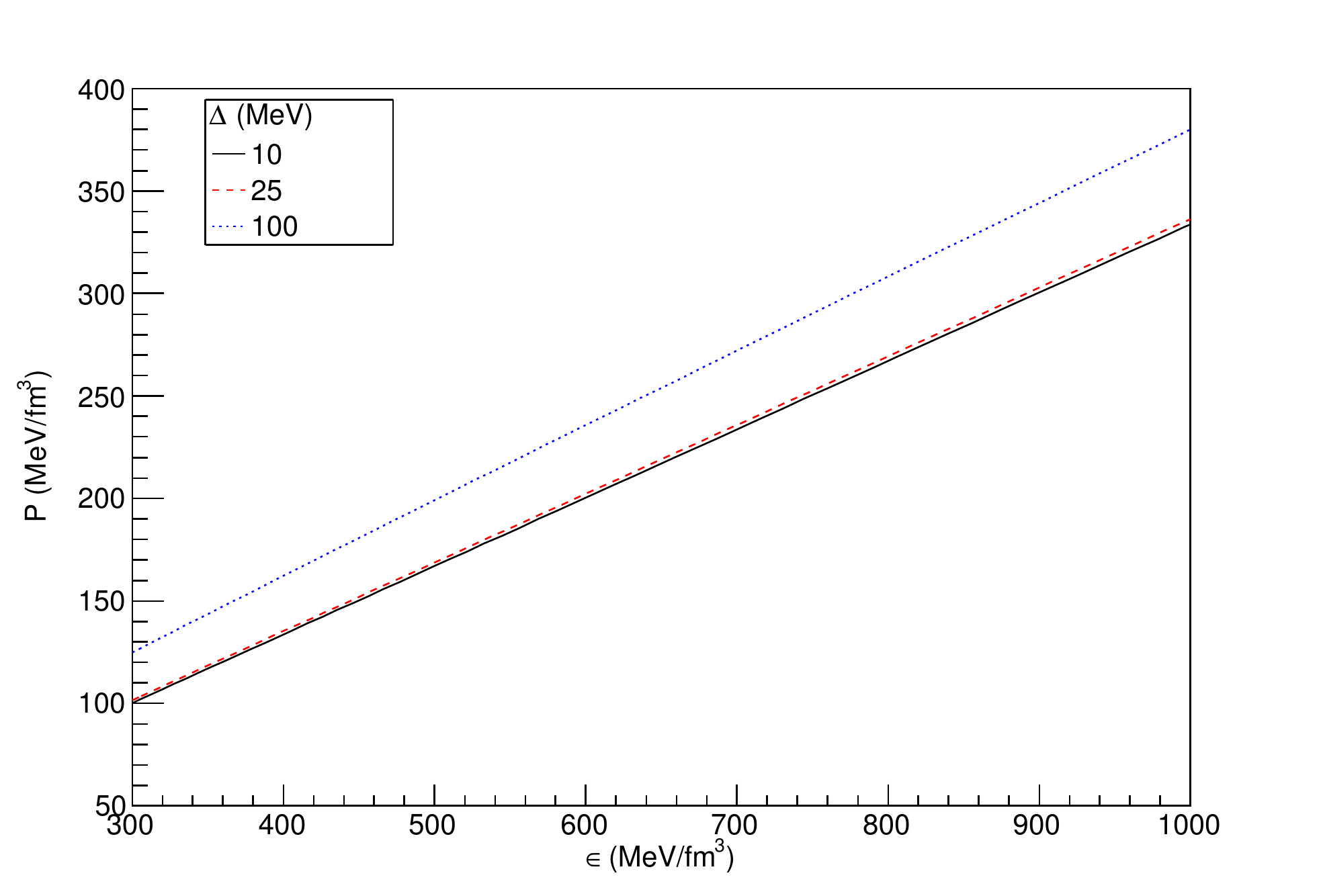}
\caption{\footnotesize The EoS for CFL matter within the NJL theory (left panel) for different values of G at $B=0$. The same for the MIT bag model for different values of $\Delta$ in the right panel.} \label{EoS}
\end{center}
\end{figure}

\begin{figure}
\begin{center}
\includegraphics[width=0.49\textwidth]{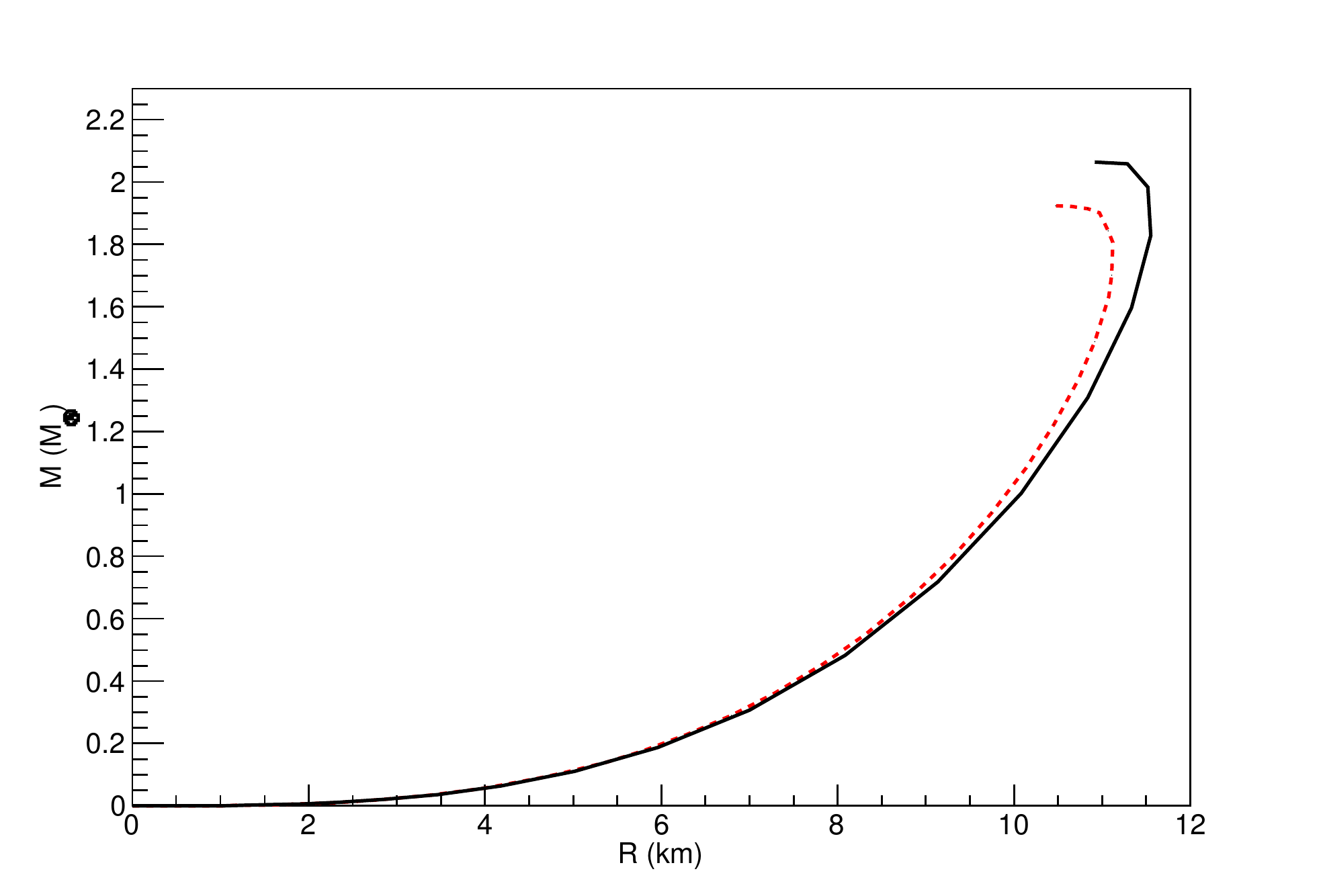}
\includegraphics[width=0.49\textwidth]{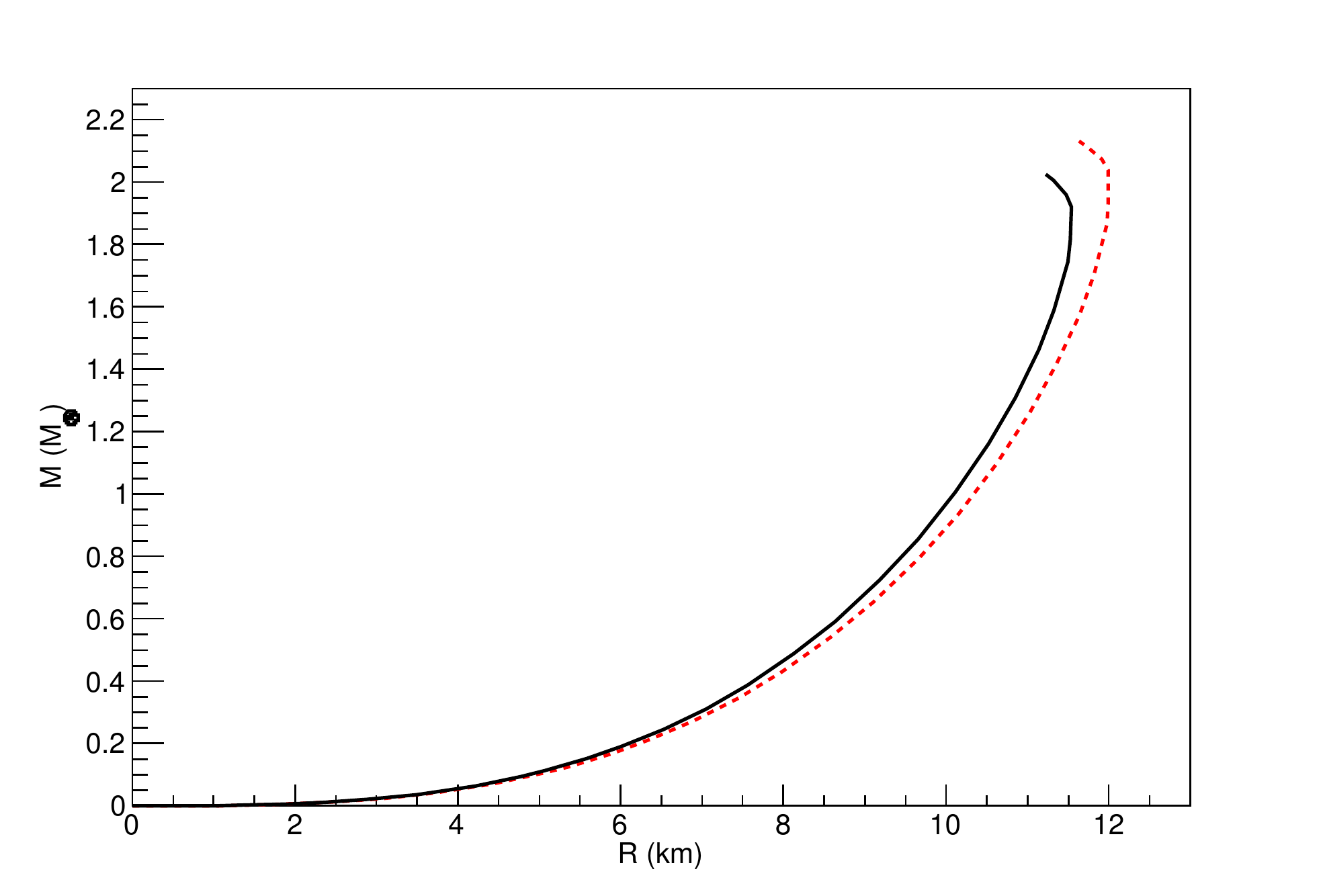}
\caption{\footnotesize Mass-radius relation for CFL matter with $B$=58 MeV/fm$^3$. Results obtained with NJL theory in the left panel: full line for G=4.32 GeV$^{-2}$ and dashed line for G=7.10 GeV$^{-2}$. Results obtained with MIT bag model in the right panel with $\Delta$ = 10 (full line) and 100 MeV (dashed line).} \label{M-R-1}
\end{center}
\end{figure}

The EoS corresponding to NJL and MIT models are represented in the region of interest for SQM in Fig. \ref{EoS} for two different values of the gap parameter. It can be observed that the splitting between the EoS for different $\Delta$'s is more significant for the values introduced in the MIT model than in those obtained in the NJL model, although they are really significant only for high gap values.
As it was found in \cite{Incera}, in the  MIT case, the higher the gap, the stiffer the EoS. Hence, as the mass supported by a given star configuration is related to the stiffness of the EoS, a higher value of the gap renders higher maximum masses for stable strange stars \cite{German}. However, this is {\it not} the case for the NJL calculations. When a higher
value of G is used, although the corresponding gap parameter increases for a fixed value of $\mu$, the EoS does not change considerably and actually weakly softens in the region of interest for neutron star interior. Therefore, in the NJL approach it is not possible to increase the maximum star mass of strange stars even when one uses unphysically large values of the coupling constant as can be seen from Fig. \ref{M-R-1}. The origin of the softening of the EoS in the CFL-NJL approach is due to the term $3\Delta^2/G$ that enters with a negative sign in the pressure. 

As discussed in  \cite{Incera}, from a physical point of view, the increase of $G$ beyond a certain value in the CFL-NJL model implies a softening of the EoS because for large enough $G$'s the system begins to crossover from  BCS to BEC \cite{BCS-BEC1}-\cite{ Israel}. The crossover is reflected in the decay of the system pressure, which is due to an increment in the number of diquarks that become Bose molecules and hence cannot contribute to the Pauli pressure of the system. As shown in Refs. \cite{Israel, Jason}, if the diquark coupling were high enough to produce a complete crossover from the BCS to the BEC regime, the pressure of the bosonic system at zero temperature would become zero implying a collapse of the stellar system.

 To conclude this section, using Eq. (\ref{Omega_MIT}) one can indeed increase the maximum allowed mass for CFL strange stars by using an arbitrarily high gap value to match the recent reported data
in \cite{Ozels1}-\cite{Demorest}.
Nevertheless, this effect is not possible within the more consistent NJL approach (\ref{OmegaCFL}), as can be seen in Fig. \ref{M-R-1}, since to increase the gap the coupling constant should be increased and it tends to decrease the mass for a given radius.

\section{EoS of the CFL-gluonic phase}

In the NJL approach, gluons degrees of freedom are usually disregarded as having a negligible effect at zero temperature. Nevertheless, as it was shown in Ref. \cite{Gluons}, gluons can affect the EoS of the CFL phase. The reason is that, as shown in \cite{Rischke2000}, the polarization effects of quarks in the CFL background embed all the gluons with Debye ($m_D$) and Meissner ($m_M$) masses that depend on the baryon chemical potential, 
\begin{equation}\label{Masses}
m_D^2=\frac{21-8\ln 2}{18} m_g^2,\;\;\;\;\;\; m_M^2=\frac{21-8\ln 2}{54}m_g^2,\;\;\;\;\;\;m_g^2=g^2\mu^2N_f/6\pi^2.
\end{equation}
Here, $N_f$ is the number of flavors of massless quarks, and $g$ the quark-gluon gauge coupling constant. As a consequence, the gluons in the CFL phase have nonzero rest energy that produces a positive contribution to the system's energy density and consequently a negative contribution to the pressure. 

To consider the gluon effect into the system EoS we started in \cite{Gluons} from the three-flavor gauged-NJL Lagrangian at finite baryon density, which is invariant under $SU_c(3)\times SU_L(3)\times SU_R(3)\times U_V(1)$ symmetry
\begin{eqnarray} \label{Lagrangian}
  \mathcal{L}=-\bar{\psi}(\gamma^\mu D_\mu+\mu\gamma^0)\psi-G_V(\bar{\psi}\gamma_\mu\psi)^2+G_S\sum_{k=0}^8\left [ \left (\bar{\psi} \lambda_k \psi\right )^2 +\left (\bar{\psi}i\gamma_5\lambda_k \psi \right )^2 \right ]  \qquad\nonumber
\\
-K\left [det_f \left ( \bar{\psi}(1 +\gamma_5)\psi \right ) +det_f \left ( \bar{\psi}(1 +\gamma_5)\psi \right )  \right ]
  +\frac{G_D}{4}\sum_\eta(\bar{\psi}P_\eta\bar{\psi}^T)(\psi^TP_\eta\psi)+\mathcal{L}_G,
\end{eqnarray}
In (\ref{Lagrangian}), the quark fields $\psi_i^a$ have flavor ($i={u,d,s}$) and color ($a={r,g,b}$) indexes. Here, the coupling $G_S$ is associated with the quark-antiquark channel, the coupling $K$ with the 't Hooft determinant term that excludes the $U_A(1)$ symmetry \cite{'tHooft}, the coupling $G_D$ with the diquark channel $P_\eta=C\gamma_5\epsilon^{ab\eta}\epsilon_{ij\eta}$ and the coupling $G_V$ with the repulsive vector channel. This last interaction term naturally appears after a Fierz transformation of a point-like four-fermion interaction with the Lorentz symmetry broken by the finite density \cite{Buballa}. We are neglecting the current quark masses $m_{0i}$ since our main domain of interest will be the CFL phase with $ \mu\gg m_{0i}$.   
The gluon Lagrangian density $\mathcal{L}_G$ is
\begin{equation} \label{Lagrangian-g} 
  \mathcal{L}_G=-\frac{1}{4}G_{\mu\nu}^AG^{\mu\nu}_A +\mathcal{L}_{gauge}+\mathcal{L}_{ghost},
\end{equation}
with $G_{\mu\nu}^A=\partial_\mu G^A_\nu-\partial_\nu G^A_\mu +gf^{ABC}G^{B}_\mu G^{C}_\nu$, the gluon strength tensor;
$\mathcal{L}_{gauge}$, a gauge-fixing term; and $\mathcal{L}_{ghost}=-\eta^{A\dagger}\partial^{\mu}(\partial_{\mu}\eta^{A}+gf^{ABC}G^{B}_{\mu}\eta^{C})$, the ghost-field Lagrangian.
The coupling between gluons and quarks occurs through the covariant derivative 
\begin{equation} \label{CovDerivative}
D_\mu= \partial_\mu -i gT^AG^A_\mu,
\end{equation}
where $T^A=\lambda^A/2$, $A=1-8$,  are the generators of the $SU_c(3)$ group in the fundamental representation ($\lambda^A$ are the Gell-Mann matrices). The model (\ref{Lagrangian}) is nonrenormalizable due to the four-fermion interaction terms, so a cutoff $\Lambda$ needs to be introduced in all the calculations to regularize the theory in the ultraviolet region. The parameter $\Lambda$ defines the energy scale below which this effective theory is valid. 

We want to consider the region of densities large enough to have a CFL phase, where the chiral condensate has already vanished and hence only the expectation values for the diquark condensate $\Delta_\eta=\langle\psi^TP_\eta\psi\rangle$ and the baryon charge density $\rho=\langle\bar{\psi}\gamma_0\psi\rangle$ should be considered. One can now bosonize the four-fermion interaction via a Hubbard-Stratonovich transformation and then take the mean-field approximation to obtain the zero-temperature thermodynamic potential \cite{Gluons},
\begin{equation}\label{modelo}
\Omega=\Omega_{q}+ \Omega_{g}  - \Omega_{vac},
\end{equation}
where the quark, gluon and vacuum contributions are given respectively by
\begin{equation} \label{Gamma0}
\Omega_{q}= -\frac{1}{4\pi^2}\int_0^\Lambda dp p^2 (16|
\epsilon|+16|\overline{\epsilon}|)-\frac{1}{4\pi^2}\int_0^\Lambda
dp p^2 (2|\epsilon'|+2|\overline{\epsilon'}|) +\frac{3\Delta^2}{G_D}-G_V\rho^2 ,
\end{equation}

\begin{equation}\label{TP-gluons-T0}
 \Omega_{g}=\frac{2}{\pi^2}\int_0^\Lambda dp p^2\left (\sqrt{p^2+ \tilde{m}^2_D \theta({\Delta}-p)+3\tilde{m}^2_g\theta(\tilde{\mu}-p)\theta(p-\Delta)}+3\sqrt{p^2+\tilde{m}_M^2\theta(\Delta-p)}\right ),
 \end{equation}
 
 \begin{equation}\label{TP-vacuum0}
\Omega_{vac}\equiv \Omega(\mu=0, \Delta=0),
 \end{equation}
 
 with energy spectra
\begin{equation}\label{Spectra}
\varepsilon=\pm \sqrt{(p-\tilde{\mu})^2+\Delta^2}, \quad
\overline{\varepsilon}=\pm \sqrt{(p+\tilde{\mu})^2+\Delta^2},\nonumber
\end{equation}
\begin{equation}\label{Spectra-2}
\varepsilon'=\pm \sqrt{(p-\tilde{\mu})^2+4\Delta^2,}\quad
 \overline{\varepsilon}'=\pm \sqrt{(p+\tilde{\mu})^2+4\Delta^2},
\end{equation}
and dynamical quantities, $\Delta$ and  
$\rho$, found from the equations,

\begin{equation} \label{Gap-Eq}
\frac{\partial\Omega}{\partial\Delta}=0,\;\;\;\;\rho=-\frac{\partial\Omega_q}{\partial\tilde{\mu}}
\end{equation}

We have that, as the mean value $\langle\bar{\psi}\gamma_0\psi\rangle$ enters in the covariant derivative as a shift to the baryon chemical potential, the effective chemical potential for the baryon charge is now $\widetilde{\mu}=\mu-2G_V \rho$ instead of $\mu$. Moreover, while the solution of the gap equation (first equation in (\ref{Gap-Eq})) is a minimum of the thermodynamic potential, the solution of the second equation is a maximum \cite{Kitazawa2002}, since it defines, as usual in statistics, the particle number density $\rho$.

To find the EoS of this system we need the corresponding pressure $P$ and energy density $\epsilon$, that can be found from the thermodynamic potential (\ref{modelo}) as
\begin{equation}\label{energy}
P= -(\Omega_{q}+ \Omega_{g}  - \Omega_{vac})+(B-B_0), \quad \epsilon = \Omega_{q}+ \Omega_{g}  - \Omega_{vac} + \tilde{\mu} \rho-(B-B_0)
\end{equation}
Notice that the chemical potential that multiplies the particle number density $\rho$ in the energy density is $\tilde{\mu}$ instead of $\mu$. This result can be derived following the same calculations of Ref. \cite{Israel-2} to find the quantum-statistical average of the energy-momentum tensor component $\tau_{00}$. 

In (\ref{energy}), we added, as usual, the bag constant B ($B_0$ is introduced to ensure that $\epsilon=P=0$ in vacuum). As we already pointed out, in the MIT bag model, $B$ was introduced as a phenomenological input parameter to account for the free energy cost of quark matter relative to the confined vacuum \cite{MIT}. Nevertheless, in the NJL model, the bag constant can be calculated in the mean-field approximation as a dynamical quantity related to the spontaneous breaking of chiral symmetry \cite{Oertel}.
Following this dynamical approach, we showed in \cite{Gluons} that the bag constant in this case is given by
$B=0$ and the only remaining bag parameter in (\ref{energy}) is $B_0$, which is written as
\begin{eqnarray}\label{Dynamical-B0}
B_0=\frac{9}{\pi^2}\left[\int_0^\Lambda
p^2dp \left (\sqrt{m^2+p^2}-\sqrt{p^2}+\frac{2G_Sm}{\sqrt{m^2+p^2}}\right ) \right ]-4K\left (\frac{3}{\pi^2} \right )^3\left [ \int_0^\Lambda dp p^2 \frac{m}{\sqrt{m^2+p^2}} \right ]^3,
\end{eqnarray}
with the same vacuum dynamical mass for the three quarks found from    
\begin{equation}
1=4G_S\frac{3}{\pi^2}\int_{0}^\Lambda p^2 dp \frac{1}{\sqrt{m^2+p^2}}+2K\frac{9}{\pi^4}\left [\int_{0}^\Lambda p^2 dp \frac{m}{\sqrt{m^2+p^2}}\right ]^2
\label{Gap-Eq-Vac}
\end{equation}
For the parameter set under consideration one obtains $B_0=57.3$ MeV/fm$^3$ \cite{Oertel}, which is a value very close to what we were using in Section 2. Nevertheless, we should underline that in this calculation no contribution associated with the confinement/deconfinement phase transition has been taken into account.

\begin{figure}
\begin{center}
\includegraphics[width=0.49\textwidth]{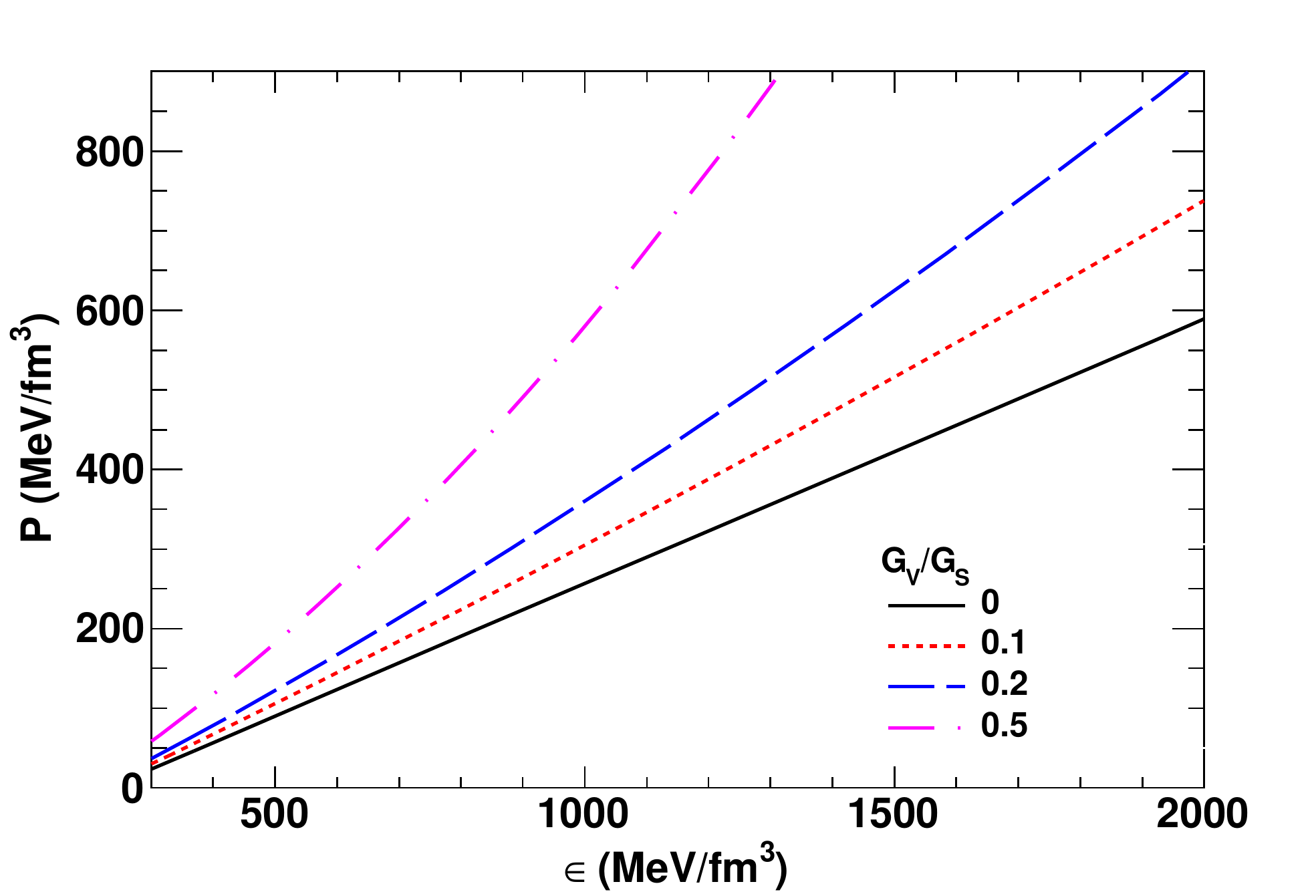}
\includegraphics[width=0.49\textwidth]{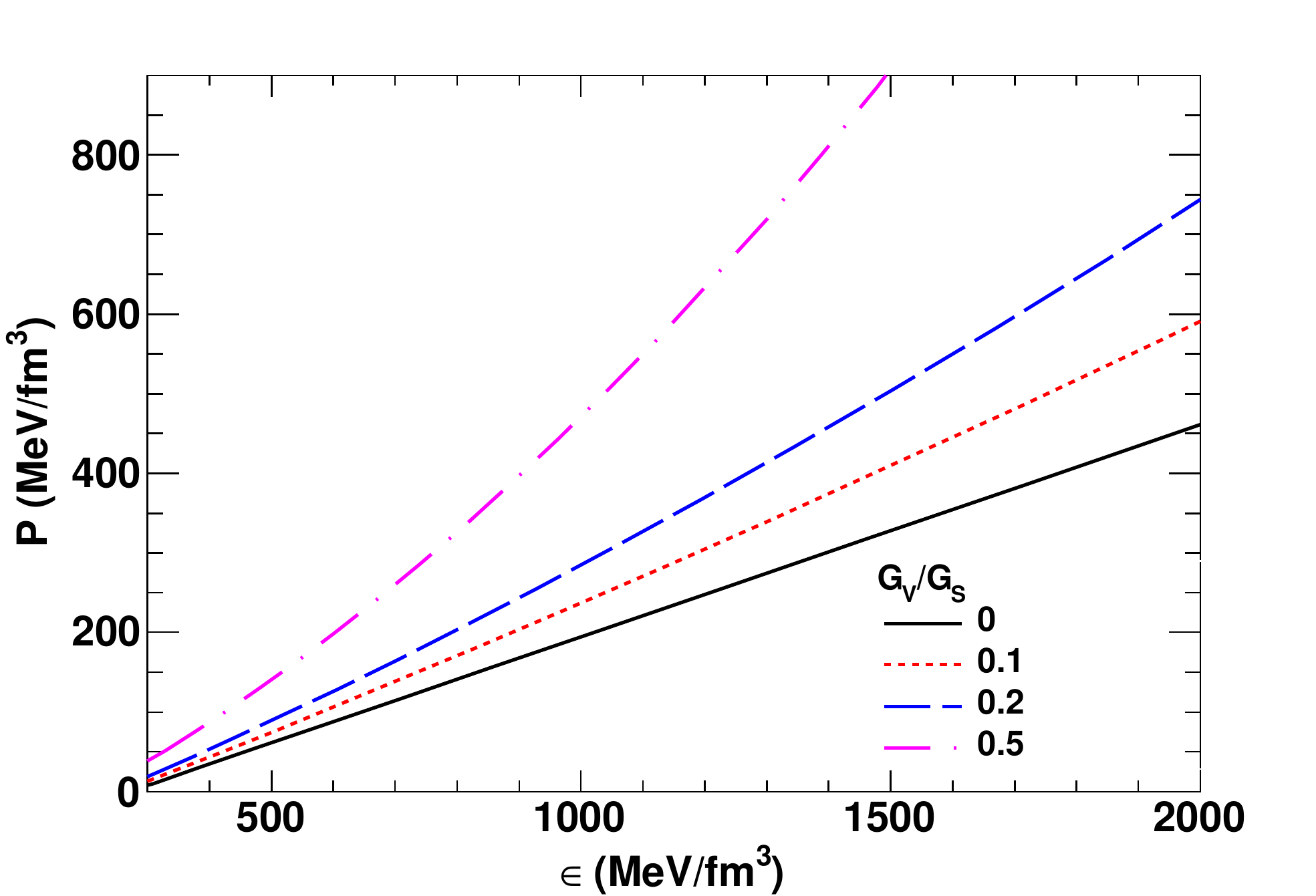}
\caption{Equation of state for CFL matter with gluon contribution (right panel) and without it (left panel), for different values of the vector coupling $G_V$.}\label{figEoS}
\end{center}
\end{figure}

Then, the EOS for a range of $G_V$ with and without the gluon contribution to the thermodynamic potential was found in \cite{Gluons} and is plotted in Fig. \ref{figEoS}. As can be seen by comparing the two panels, the inclusion of gluons' degrees of freedom soften the EOS. Only for relatively high values of $G_V$ the EOS of the CFL matter with gluons becomes stiffer than the EOS of the regular CFL matter with no gluons and no vector coupling, at least in the energy density region relevant for compact stars. Moreover, if we compare the curve for the CFL matter at certain value of $G_V$ with the corresponding one including gluons and the same coupling value, the former will be stiffer than the second one, independent of the $G_V$ value.

We can also note that with the increase of the energy density the shift in pressure due to a change in $G_V$ for CFL matter with gluons  increases for the same value of energy density. On the other hand, while at energy densities smaller than $500$ MeV/fm$^3$ there is a negligible change even for $G_V/G_S=0.5$, at $\epsilon=1500$ MeV/fm$^3$, the jump in pressure is really noticeable.  This is a natural result, since the effect of the vector interaction is proportional to the baryon density and should be more significant in the high density region. 

\begin{figure}
\begin{center}
\includegraphics[width=0.49\textwidth]{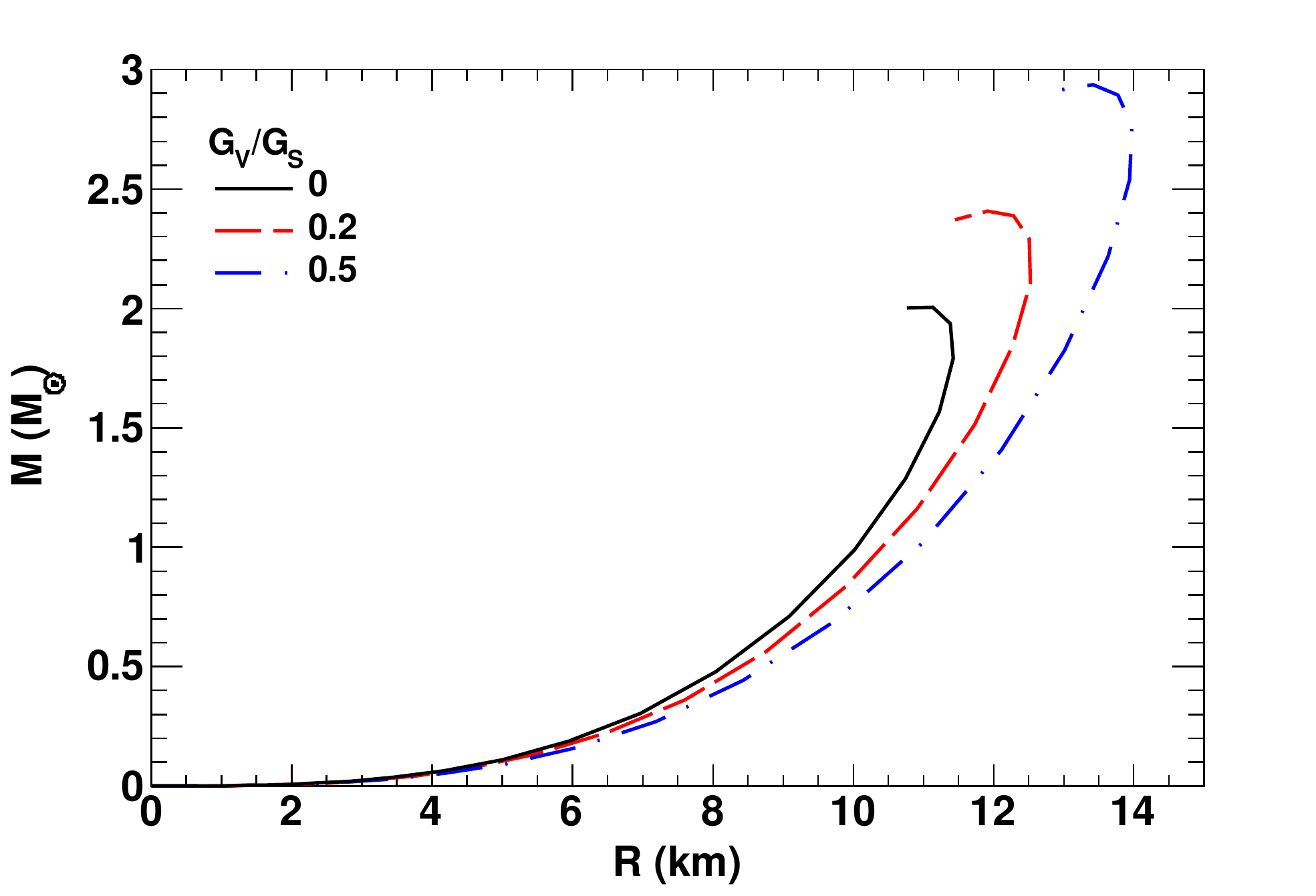}
\includegraphics[width=0.49\textwidth]{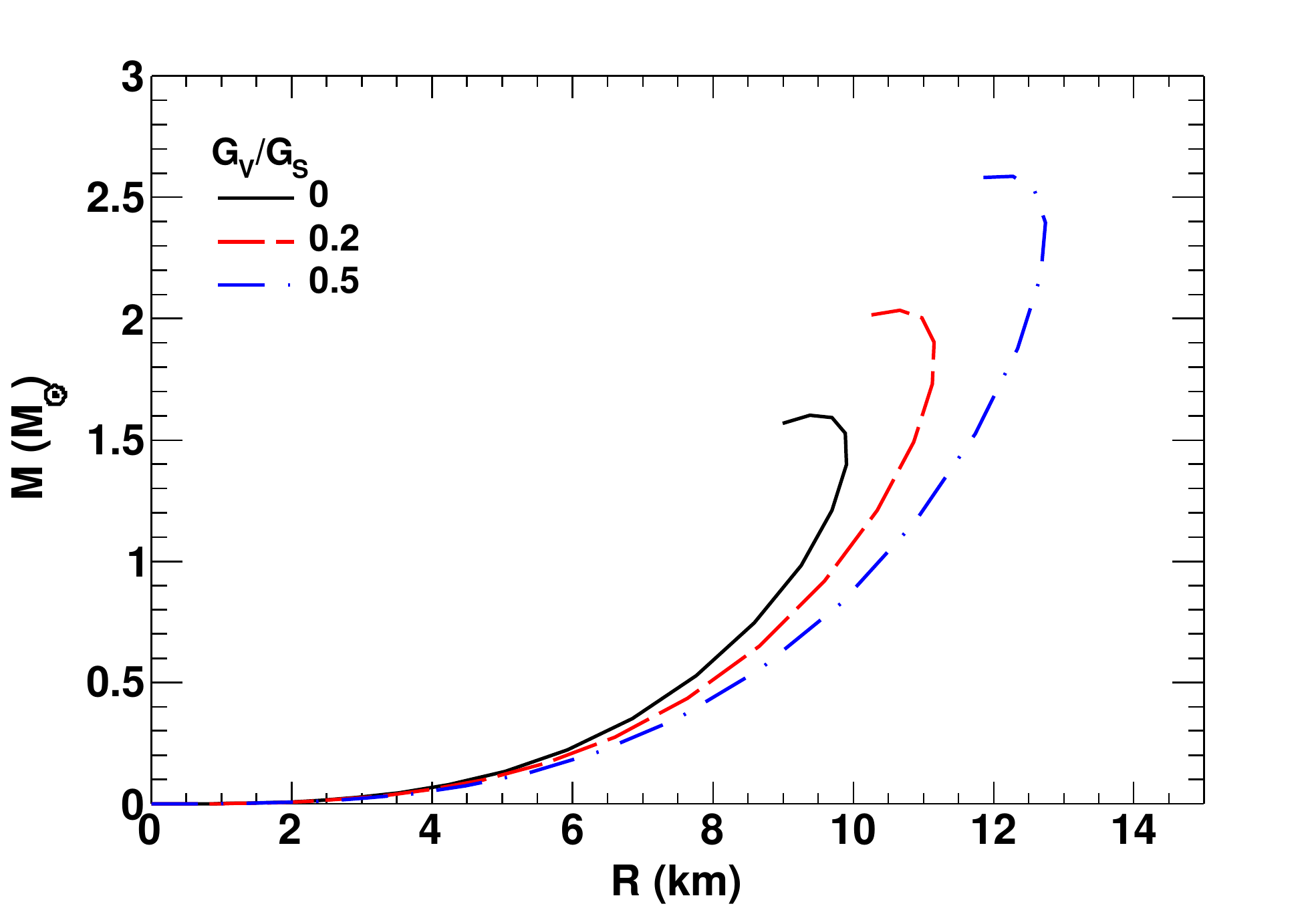}
\caption{Mass-radius relationship of strange stars made of CFL matter with (right panel) and without (left panel) gluon contribution. The maximum mass values for vector interaction in the range $G_V/G_S<0.2$ remains below 2$M_\odot$.} \label{figMR}
\end{center}
\end{figure}
Now it can be calculated the M-R relationship for strange stars made of CFL matter with gluon contribution by using Eqs. (\ref{TOV1}) and  (\ref{TOV2}). The corresponding M-R sequences found in \cite{Gluons} are shown in Fig. \ref{figMR} for the EoS of quark matter with (right panel) and without (left panel) the gluon contribution. Comparing them, it is evident that the gluons decrease the maximum mass for each sequence up to $20 \%$, an effect that turns more prominent for lower values of the vector interaction. Sequences including gluons cannot reach $2M_{\odot}$ if $G_V/G_S<0.2$. Here, it is important to point out that several results actually suggest that a low vector coupling between quarks may be favored at high-densities  \cite{GV-Vacuum}. Even more, as discussed in \cite{Kashiwa}, the vector interaction makes the chiral phase transition weaker in the low $T$ and high $\mu$ region, with the possibility that  the transition becomes a crossover in the region
when the interaction is strong enough \cite{Kitazawa2002, Abuki}, what is not expected from Lattice calculations. Then, the absence of the vector interaction would be preferable in the high density region. The same conclusion was posed by studying the phase transition from the hadronic phase to the quark phase in Ref. \cite{Bentz}. There, it was found that the hadron-quark phase transition may take place only at small $G_V/G_S$ ratio. For larger ratio, the repulsive vector interaction makes the NJL phase too stiff to cross to the hadronic phase. However, taking a zero vector interaction for quark matter and further considering effects which soften the quark matter EOS, like diquark condensation, there could be a phase transition.

Then, if it is needed to consider a very weak or even zero vector interaction for high-dense quark matter, gluon effects would rule out stars composed entirely of CFL matter. More precisely, stars composed entirely of CFL quark matter with gluons, within the 
formalism presented here, should be disregarded if $G_V$ is sufficiently low, as they cannot explain the mass measurements of the mentioned compact stars PSR J1614-2230 and PSR J0348+0432.

\section{EoS of Magnetized Quark Matter}

In the presence of a uniform magnetic field $H$, the EoS becomes anisotropic \cite{Israel-2}. In this case, the pressure of the system develops a splitting in the directions parallel and perpendicular to the applied field. This splitting has to be taken into account in the EoS, which is then modified from the usual form into 

\begin{equation}
P_{\parallel} =-\Omega-\frac{H^2}{2} \,,\\
\label{eq:pperp}
\end{equation}
\begin{equation}
P_{\perp}  =-\Omega - H\mathcal{M} + \frac{H^2}{2} \,, \\
\end{equation}
\begin{equation}
\varepsilon  =\Omega+\mu \rho +\frac{H^2}{2} \,, \label{eq:split}
\end{equation}
where $\Omega$ is the thermodynamical potential evaluated at the physical minimum, $\rho=-\partial\Omega/\partial\mu$ is the particle density, $\varepsilon$ the energy density and ${\cal M}=-\partial\Omega/\partial H$ the system magnetization.

\begin{figure}
\begin{center}
\includegraphics[width=0.47\textwidth]{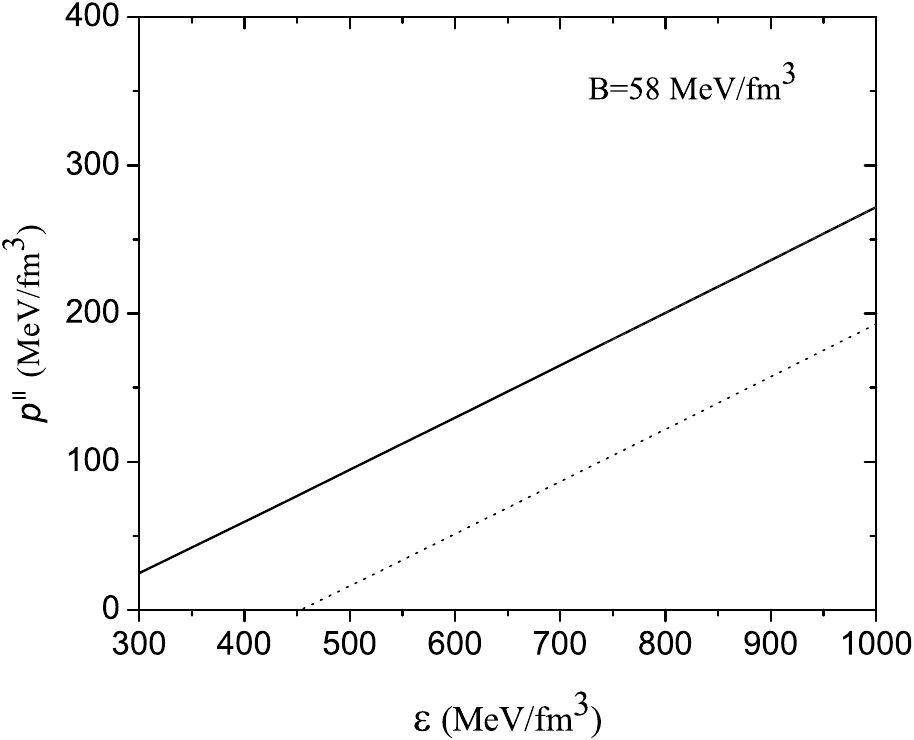}
\includegraphics[width=0.47\textwidth]{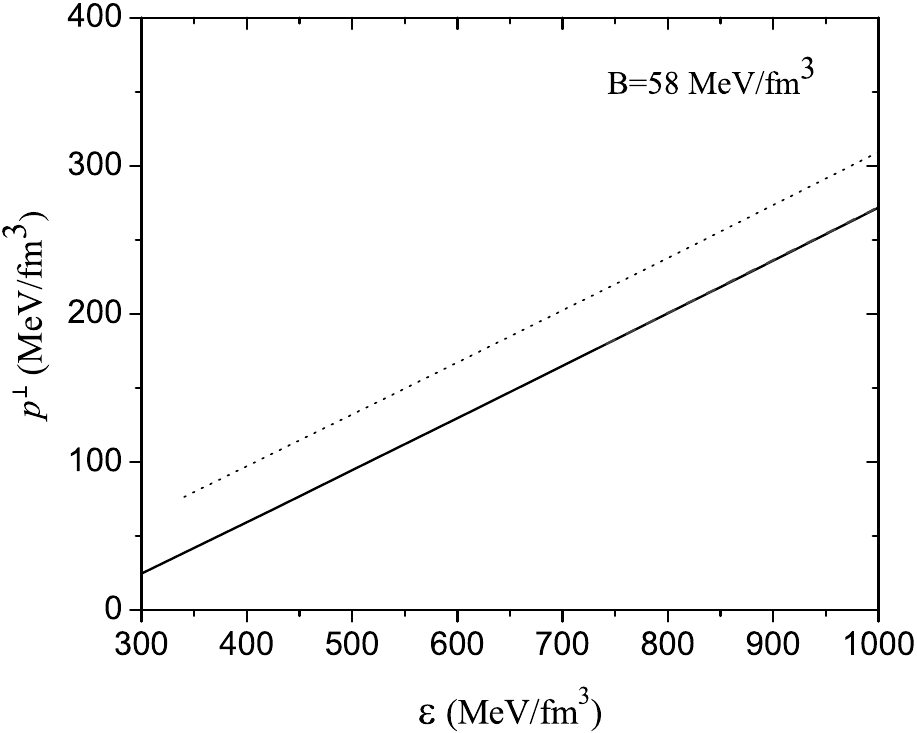}
\caption{\footnotesize EoS for MCFL matter
considering parallel (left panel) and perpendicular (right panel)
pressures for different values of $\tilde{H}$: zero field (solid line),
and $5 \times 10^{18} G$ (dotted line). The value of
the bag constant was fixed to B=58 MeV/fm$^3$.} \label{EOS}
\end{center}
\end{figure}

In the case of SQM in the color superconducting phase, the magnetic field gives rise to a new phase that is called the magnetic CFL (MCFL) phase (see for review \cite{MCFL-review} and references therein).  In this case the thermodynamic potential is given by \cite{Paulucci}
\begin{equation}
\Omega_{MCFL} =\Omega_C+\Omega_N
\label{OmegaMCFL}
\end{equation}
with

\begin{equation}\label{C}
\Omega_{C} =-\frac{\widetilde{e}\widetilde{H}}{4\pi^2}\sum_{n=0}^\infty
(1-\frac{\delta_{n0}}{2})\int_0^\infty dp_3 e^{-(p_3^2+2\widetilde{e}\widetilde{H}n)/
\Lambda^2}[8|\varepsilon^{(c)}|+8|\overline{\varepsilon}^{(c)}|],
\end{equation}

\begin{equation}
\Omega_{N} =-\frac{1}{4\pi^2}\int_0^\infty dp p^2 e^{-p^2/\Lambda^2}[6|
\varepsilon^{(0)}|+6|\overline{\varepsilon}^{(0)}|]-\frac{1}{4\pi^2}\int_0^\infty
dp p^2 e^{-p^2/\Lambda^2}\sum_{j=1}^2[2|\varepsilon_j^{(0)}|+2|\overline{\varepsilon}_j^{(0)}|]+
\frac{\Delta^2}{G}+\frac{2\Delta^2_H}{G},
\label{N}
\end{equation}
and
\begin{eqnarray}
\varepsilon^{(c)}=\pm \sqrt{(\sqrt{p_3^2+2\widetilde{e}\widetilde{H}n}-\mu)^2+\Delta_H^2},\nonumber
\\
\overline{\varepsilon}^{(c)}=\pm \sqrt{(\sqrt{p_3^2+2\widetilde{e}\widetilde{H}n}+\mu)^2+\Delta_H^2},
\end{eqnarray}

\begin{eqnarray}
\varepsilon^{(0)}=\pm \sqrt{(p-\mu)^2+\Delta^2},\qquad \overline{\varepsilon}^{(0)}=\pm \sqrt{(p+\mu)^2+\Delta^2},\nonumber
\\
\varepsilon_1^{(0)}=\pm \sqrt{(p-\mu)^2+\Delta_a^2},\qquad \overline{\varepsilon}_1^{(0)}=\pm
\sqrt{(p+\mu)^2+\Delta_a^2},\nonumber
\\
\varepsilon_2^{(0)}=\pm \sqrt{(p-\mu)^2+\Delta_b^2},\qquad \overline{\varepsilon}_2^{(0)}=\pm
\sqrt{(p+\mu)^2+\Delta_b^2},
\end{eqnarray}
being the dispersion relations of the charged $(c)$ and neutral $(0)$ quarks. In the above we used the notation
$\Delta_{a/b}^2=\frac{1}{4}(\Delta\pm\sqrt{\Delta^2+8\Delta_H^2})^2$. In this formulation we are neglecting again the quark masses since they are too much smaller than the chemical potential at the densities under consideration. Also we are not considering the possible effect of the magnetic field interaction with the quark anomalous magnetic moment, since as proved in \cite{Angel} it has a negligible effect. Even the third gap appearing in the MCFL phase, which is associated to the Cooper condensate magnetic moment \cite{Bo} is not taken into consideration because it is only relevant at extremely high magnetic fields when the particles are confined to the lowest Landau level.

The MCFL gaps $\Delta$ and $\Delta_{H}$ correspond respectively to the $(d,s)$ pairing gap, which takes place only between
neutral quarks, and to the $(u,s)$ and
$(u,d)$ pairing gaps, which receive contribution from pairs of
charged and neutral quarks. The separation of the gap in
two different parameters in the MCFL case, as compared to the CFL,
where there is only one gap, reflects the symmetry
difference between these two phases \cite{Cristina-1, Cristina-2}. Here again, $\Lambda$-dependent smooth cutoffs were introduced. The notation,  $\widetilde{H}$, is used for the in-medium rotated magnetic field that due to his long-range nature can penetrate the color superconductor \cite{Gatto}.

\begin{figure}
\begin{center}
\includegraphics[width=0.47\textwidth]{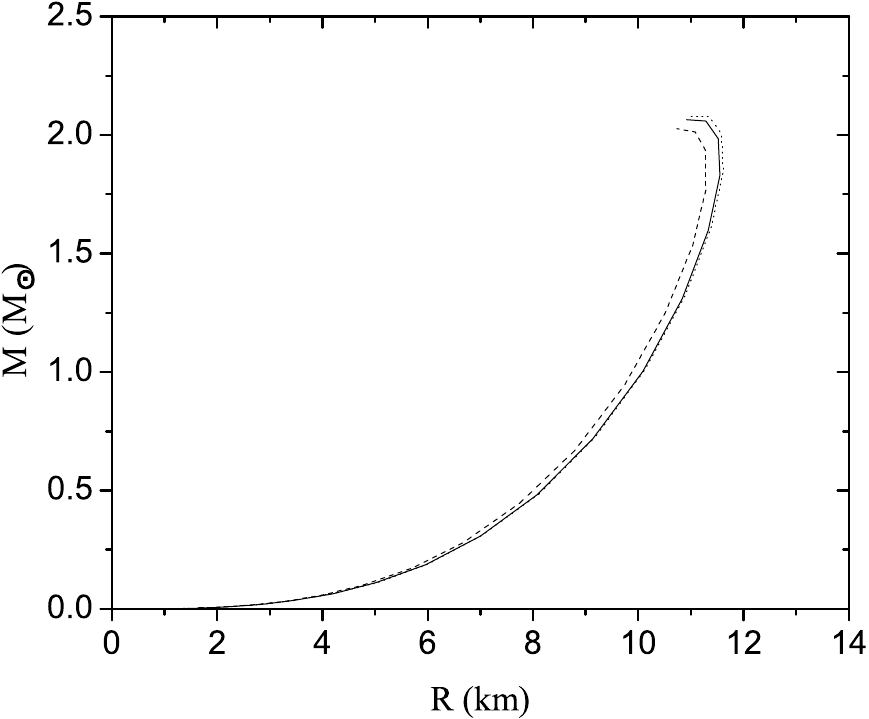}
\includegraphics[width=0.47\textwidth]{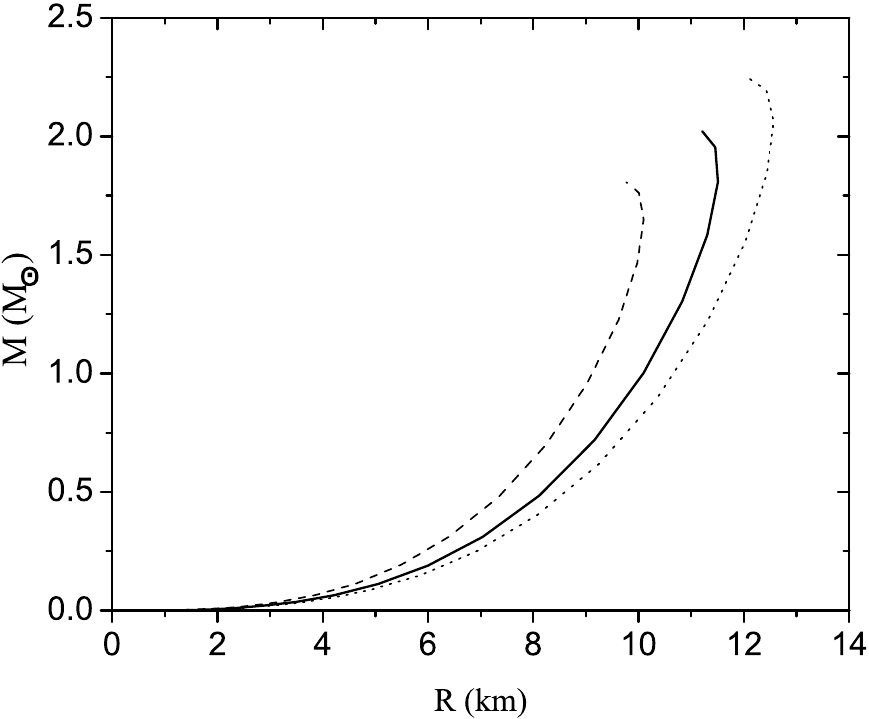}
\caption{\footnotesize Mass-radius relation for strange stars in the MCFL phase with bag constant $B=58$ MeV/fm$^3$. The full line
indicates the M-R relation for zero magnetic field, whereas the
dashed and dotted lines corresponds to MR relation calculated with
parallel and perpendicular pressures, respectively, for
$\tilde{H}=1.7\times 10^{17}$ G (left panel) and
$\tilde{H}=3\times 10^{18}$ G (right panel).} \label{MR}
\end{center}
\end{figure}

The EoS, corresponding to Eqs. (\ref{eq:pperp})-(\ref{eq:split}) evaluated in the solutions of the gap equations 
\begin{equation}
\frac{\partial \Omega_{MCFL}}{\partial \Delta}=0,\qquad\qquad \frac{\partial \Omega_{MCFL}}{\partial \Delta_H}=0,
\label{Delta-MCFL}
\end{equation}
are plotted in Fig. \ref{EOS}. There, we can see that in the presence of a magnetic field the EoS becomes highly anisotropic. While for the parallel pressure the magnetic field shifts the EoS curve above the zero-field one, for the parallel pressure the magnetic field effect is the opposite. Moreover, the magnetic-field effect is much significant for the
$\epsilon-p^\|$ relationship than for the $\epsilon-p^\bot$ one. 
At $\widetilde{H}\sim 10^{18}$ G, the
shift in the energy density with respect to the zero-field value is
$\sim 200$ MeV/fm$^3$ for the same pressure  in the parallel-pressure case, while this field effect
is too much smaller for the same
range of field values in the perpendicular-pressure case.

Now we can use the obtained EoS to obtain the M-R relationship through the TOV equations (\ref{TOV1})-(\ref{TOV2}) for the parallel and perpendicular pressures independently. The results are plotted in Fig. \ref{MR}. There we can see that the effect of the parallel pressure is to increase the star mass for a given radius, while the perpendicular pressure tends to decrease the mass for the same radius. Hence, the anisotropy of the magnetized EoS opens an interesting question in regards to the calculation of the star M-R relationship, since each pressure has an opposite effect. To solve this paradox it will be needed to find an alternative formulation that includes the anisotropy from where to obtain the M-R relationship, since the TOV equations (\ref{TOV1})-(\ref{TOV2}) were obtained under the supposition that the system has an isotropic configuration.

\section {Summary}

In this paper it was reviewed the EoS's of quark matter under several conditions and following different approaches with the goal to see under what conditions the observed mass value M = 1.97$\pm$0.04 $M_{\odot}$ can be reached. 

First, the results for the EoS in the MIT bag model was compared to those of the NJL model. It was shown that while in the MIT model the introduction by hand of a large gap produces a stiffer EoS, when in the NJL model it is increased the gap self consistently by strengthening the diquark-diquark coupling the EoS becomes softer. This last effect is associated with the fact that a strong coupling can produce a BCS-BEC crossover with a decrease of pressure in the BEC phase. This result implies that to use the MIT bag model with large gap values to predict stellar masses can be misleading. Also, it is worth to notice that the maximum value of the mass that can be reached in both models is sensitive to the used bag constant value.

Then, it was consider the effect of gluons in the EoS of dense quark matter in the CFL phase at zero temperature. Since in this phase the gluons acquire Debye and Meissner masses they can contribute to the thermodynamic potential calculated in the framework of the gauged-NJL model. The gluon masses increase the system rest energy at zero temperature and hence decrease the pressure. As a consequence, the gluon effect softer the EoS in such a way that only for values of the coupling constant associated with the vector-vector interaction, $G_V$,  larger than a certain critical value, the system can reach the observed stellar mass value  M = 1.97$\pm$0.04 $M_{\odot}$.

Also, the effect of a strong magnetic field on the EoS and M-R relation for CFL matter was discussed. The main result is that a sufficiently high magnetic field produces a big anisotropy in the EoS that prevents
the use of the TOV equations to study the M-R relationship in those conditions, since the TOV equations were derived for an isotropic medium with spherical symmetry. 
Moreover, it is known that at sufficiently high $\widetilde{H}$ new phases of dense quark matter can take place \cite{phase}, as for example, a paramagnetic phase with a ground state of gluon vortices that are formed at rotated magnetic fields higher than the gluon Meissner mass square \cite{vortices}-\cite{vortices-2}. How this magnetic phase can affect the EoS of dense quark matter is an open question.

Finally, to extrapolate the results presented here to hybrid stars, other quark phases that are more likely to be realized at the moderated-high densities that can prevails in the core  close to the transition from the existing nuclear matter in the envelope, should be considered. Other phases of quark matter,  as 2SC at strong coupling with \cite{2SC-B} or without \cite{2SC} magnetic field, or the magnetized neutral DCDW phase \cite{PRD92}, are good candidates in this context.


\begin{acknowledgments}

I would like to thank my collaborators V. de la Incera, J. E. Horvath and L. Paulucci who contributed to must of the results reviewed here.

\end{acknowledgments}


\section{References}

\begin{thebibliography}{15}

\bibitem{Bodmer}  Bodmer A 1971 \textit{Phys. Rev. D} \textbf{4} 1601

\bibitem{Chin}  Chin S A and Kerman A K 1979 \textit{Phys. Rev. Lett.} \textbf{43} 1292


\bibitem{Terazawa}  Terazawa H 1979 \textit{Tokyo U. Report} INS-336


\bibitem{Wit}  Witten E 1984 \textit{Phys. Rev. D} \textbf{30} 272
 

 \bibitem{Pairing1} Alford M, Rajagopal K and Wilczek F 1999  \textit{Nucl. Phys. B} \textbf{537} 433
  
  
 \bibitem{Pairing2} Rapp R, Schaefer T, Shuryak E V and Velkovsky M 2000  \textit{Ann. Phys. (N.Y.)} \textbf{280} 35
 
 \bibitem{Pairing3} Alford M, Rajagopal K, Reddy S and Wilczek F 2001  \textit{Phys. Rev. D} \textbf{64} 074017

 \bibitem{German} Lugones G and Horvath J E 2002   \textit{Phys. Rev. D} \textbf{66} 074017
 
    
\bibitem{Ozels1}  Ozel F and Psaltis D 2009  \textit{Phys. Rev. D} \textbf{80} 103003

\bibitem{Ozels2} Ozel F, Baym G and Guver T 2010 \textit{Phys. Rev. D} \textbf{82} 101301

 \bibitem{Demorest} Demorest P B et al. 2010   \textit{Nature} \textbf{467} 1081

\bibitem{Lattimer} Steiner A W, Lattimer J M and Brown E F 2010 \textit{Astrophys. J.} \textbf{722} 33   

 
  \bibitem{MIT} Chodos A, Jaffe R L, Johnson K, Thorn C B and Weisskopf V F 1974  \textit{Phys. Rev. D} \textbf{9} 3471
   
   
   \bibitem{Alford04}  Alford M, Kouvaris C and Rajagopal K 2004 \textit{Phys. Rev. D} \textbf{92} 222001

 \bibitem{Rajagopal2001} Rajagopal K and Wilczek F 2001  \textit{Phys. Rev. Lett.} \textbf{86} 3492
  

\bibitem{Alford2001} Alford M, Rajagopal K, Reddy S and Wilczek F 2001 \textit{Phys. Rev. D} \textbf{64} 074017


\bibitem{Paulucci} Paulucci L, Ferrer E J,  {de la Incera} V and Horvath J E 2011  \textit{Phys. Rev. D} \textbf{83} 043009

\bibitem{Incera} Paulucci L, Ferrer E J, Horvath J E and  {de la Incera} V 2013 \textit{J. Phys. G} \textbf{40} 125202


\bibitem{BCS-BEC1} Nishida Y and Abuki H 2005  \textit{Phys. Rev. D} \textbf{72} 096004

\bibitem{BCS-BEC2} Sun G-F, He L and Zhuang P 2007  \textit{Phys. Rev. D} \textbf{75} 096004

\bibitem{BCS-BEC3} He L and Zhuang P 2007 \textit{Phys. Rev. D} \textbf{75} 096003

\bibitem{Israel}Ferrer E J, de la Incera V, Keith J P and Portillo I 2015 \textit{Nucl. Phys. A}. \textbf{933} 229

\bibitem{Jason} Ferrer E J and Keith J P 2012 \textit{Phys. Rev. C} \textbf{86} 035205

\bibitem{Gluons} Ferrer E J, de la Incera V and Paulucci L 2015 \textit{Phys. Rev. D} \textbf{92} 043010

\bibitem{Rischke2000} Rischke D H 2000 P\textit{hys. Rev. D} \textbf{62} 054017 

\bibitem{'tHooft} Hooft 't  1976 \textit{Phys. Rev. Lett.} \textbf{37} 8  

\bibitem{Buballa} Buballa M 2005  \textit{Phys. Rep.} \textbf{407} 205 


\bibitem{Kitazawa2002} Kitazawa M, Koide T, Kunihiro T and Nemoto Y 2002 \textit{Prog. Theor. Phys.} \textbf{108} 929 

\bibitem{Israel-2} Ferrer E J, de la Incera V, Keith J P, Portillo I and Springsteen P L 2010 \textit{Phys. Rev. C} \textbf{82} 065802 


\bibitem{Oertel} Buballa M and Oertel M 1999 \textit{Phys. Lett. B} \textbf{457} 261

\bibitem{GV-Vacuum} Steinheimer J and Schramm S 2014 \textit{Phys. Lett. B} \textbf{736} 241 

\bibitem{Kashiwa} Kashiwa K, Kouno H, Sakaguchi T, Matsuzaki M and Yahiro M 2007 \textit{Phys. Lett. B} \textbf{647} 446 
 
  \bibitem{Abuki} Abuki H, Gatto R and Ruggieri M 2009 \textit{Phys. Rev. D} \textbf{80} 074019 
 
 \bibitem{Bentz} Bentz W, Horikawa T, Ishii N and Thomas A W 2003 \textit{Nucl. Phys. A} \textbf{720} 95 
 
 \bibitem{MCFL-review} Ferrer E J and de la Incera V 2013 \textit{Magnetism in Dense Quark Matter}   (Lect. Notes in Phys. vol.  871 )  ed. by Kharzeev D, Landsteiner K, Schmitt A andYee  H.-U.  (Springer Berlin Heidelberg) pp. 399-432
 
\bibitem{Angel} Ferrer E J, de la Incera V, Manreza Paret D, Perez Martinez A and Sanchez A 2015 \textit{Phys. Rev. D} \textbf{91} 085041

\bibitem{Bo} Feng B, Ferrer E J and de la Incera V 2011 \textit{Nucl . Phys. B} \textbf{853} 213

 \bibitem{Cristina-1} Ferrer E J, de la Incera V and Manuel C 2005 \textit{Phys. Rev. Lett.} \textbf{95} 152002

\bibitem{Cristina-2} Ferrer E J, de la Incera V and Manuel C 2006 \textit{Nucl. Phys. B} \textbf{747} 88

\bibitem{Gatto} Alford M, Berges J and Rajagopal K 2000 \textit{Nucl. Phys. B} {\bf 571} 269 

\bibitem{phase} Ferrer E J and de la Incera V 2007 \textit{Phys. Rev. D} \textbf{76} 045011

\bibitem{vortices} Ferrer E J and de la Incera V 2006  \textit{Phys. Rev. Lett} \textbf{97} 122301

\bibitem{vortices-2} Ferrer E J and de la Incera V 2007  \textit{J. Phys. A} \textbf{40} 6913

\bibitem{2SC-B} Mandal T and Jaikumar P 2016  \textit{Phys.Rev. D} \textbf{94} 074016


\bibitem{2SC}  R\"{u}ster S B, Werth V, Buballa M, Shovkovy I A and Rischke D H 2005 \textit{Phys. Rev. D} \textbf{72} 034004



\bibitem{PRD92} Carignano S, Ferrer E J, de la Incera V and Paulucci L 2015 \textit{Phys. Rev. D} \textbf{92} 105018

 
 \end{thebibliography}
\end{document}